\documentclass[sigconf]{acmart}
\usepackage{subcaption}
\usepackage{algorithm}
\usepackage{algorithmic}

\usepackage{blindtext}
\AtBeginDocument{%

    }
\setcopyright{acmlicensed}
\copyrightyear{2025}
\acmYear{2025}
\acmDOI{XXXXXXX.XXXXXXX}
\acmConference[HPDC'25]{In The 34th International Symposium on High Performance Parallel and Distributed Computing }{July 20--23, 2025}{Notre Dame, IN, USA}
\acmISBN{978-1-4503-XXXX-X/18/06}

\begin{document}
\begin{sloppypar}
\title{SpeedLLM: An FPGA Co-design of Large Language Model Inference Accelerator}

%%\author{Peipei Wang}
%\email{wangpeipei@bupt.edu.cn}
%\email{webmaster@marysville-ohio.com}
%\affiliation{%
%  \institution{Institute for Clarity in Documentation}
%  \country{USA}
%}

\author{Peipei Wang}
\authornote{Student author}
\affiliation{%
  \institution{Beijing University of Posts and Telecommunications}
  \state{Beijing}
  \country{China}}
\email{wangpeipei@bupt.edu.cn}
\orcid{0009-0003-7267-9366}

\author{Wu Guan}
\affiliation{%
 \institution{Beijing University of Posts and Telecommunications}
  \state{Beijing}
  \country{China}}
\email{guanwu@bupt.edu.cn}

\author{Liping Liang}
\affiliation{%
  \institution{Beijing University of Posts and Telecommunications}
  \state{Beijing}
  \country{China}}
\email{liangliping@bupt.edu.cn}

\author{Zhijun Wang}
\affiliation{%
  \institution{Beijing University of Posts and Telecommunications}
  \state{Beijing}
  \country{China}}
\email{wangzhijun@bupt.edu.cn}

\author{Hanqing Luo}
\affiliation{%
\institution{Beijing University of Posts and Telecommunications}
\state{Beijing}
\country{China}}
\email{luohanqing@bupt.edu.cn}

\author{Zhibin Zhang}
\authornote{Corresponding author}
\affiliation{%
\institution{Institute of Computing Technology, Chinese Academy of Sciences}
\state{Beijing}
\country{China}}
\email{zhangzhibin@ict.ac.cn}

\begin{abstract}
%This paper introduces SpeedLLM, a neural network accelerator designed on the Xilinx Alevo U280 platform and optimized for the Tinyllama framework to enhance edge computing performance. Key innovations include data stream parallelism, a memory reuse strategy, and Llama2 operator fusion, which collectively reduce latency and energy consumption. SpeedLLM’s data pipeline architecture optimizes the read-compute-write cycle, while the memory strategy minimizes FPGA resource demands. The operator fusion boosts computational density and throughput. Results show SpeedLLM outperforms traditional Tinyllama implementations, achieving up to 4.8× faster performance and 1.18× lower energy consumption, offering significant improvements for deploying deep learning in edge devices.
This paper introduces SpeedLLM, a neural network accelerator designed on the Xilinx Alevo U280 platform and optimized for the Tinyllama framework to enhance edge computing performance. Key innovations include data stream parallelism, a memory reuse strategy, and Llama2 operator fusion, which collectively reduce latency and energy consumption. SpeedLLM’s data pipeline architecture optimizes the read-compute-write cycle, while the memory strategy minimizes FPGA resource demands. The operator fusion boosts computational density and throughput. Results show SpeedLLM outperforms traditional Tinyllama implementations, achieving up to 4.8× faster performance and 1.18× lower energy consumption, offering improvements in edge devices.
\end{abstract}

\keywords{Large Language Model(LLM), Xilinx Alevo U280, FPGA, Neural Network Accelerator, Edge Computing}
\maketitle
\section{INTRODUCTION}
The advancements in Artificial Intelligence have ushered in an era dominated by Large Language Models (LLMs) such as GPT-4.0, Jurassic-1 and BERT Turbo. These models not only understand but respond to user inputs with remarkable precision, revolutionizing numerous sectors including customer support, real-time interaction applications, and automated content creation. Their ability to process and generate language-based data accurately appears almost limitless, playing critical roles in scenarios demanding high-performance responses — such as code completion and real-time chat functionalities[1].\par
\hspace{2em}Tinyllama represents a compressed, optimized version of larger language models designed specifically to maintain high levels of accuracy while significantly reducing the model's size and computational needs. When deployed on the scene, for example, edge servers, IoT devices, satellite communications, the architecture of Tinyllama needs to be accelerated — particularly its ability to handle diverse data and computation efficiently, which aims to address the critical balance between performance and resource usage, reducing costs and energy consumption when deploying AI applications in scale.\par
\begin{figure}[h]
	\centering
	\includegraphics[width=0.9\linewidth]{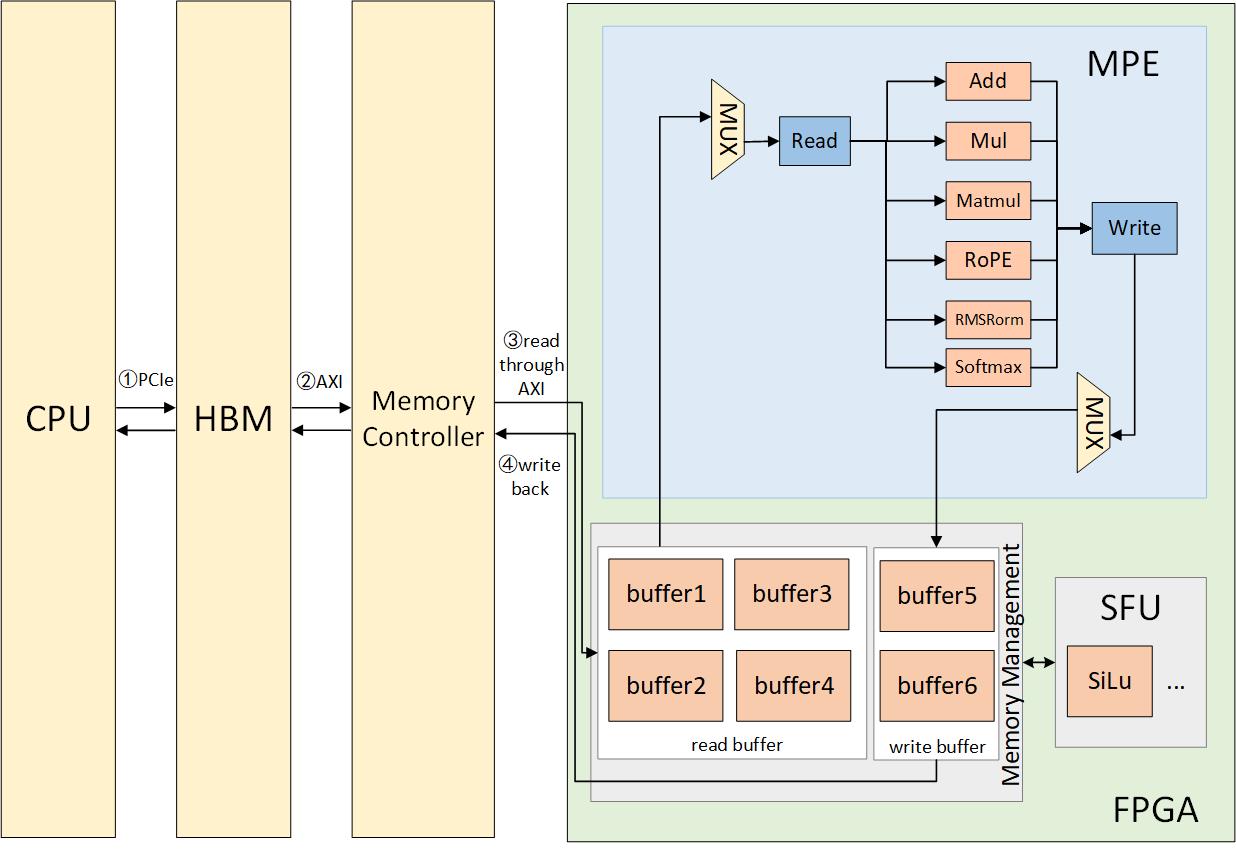}
	\caption{The overall architecture of SpeedLLM, including
Matrix Processing Engine(MPE), Memory Management, and Special Function Unit(SFU).}
	\label{fig-MultiTask}
\end{figure}
\hspace{2em}Despite their capabilities, LLMs present significant challenges primarily due to their enormous size and computational demands. For instance, models like GPT-4 may contain hundreds of billions of parameters, requiring extensive memory and computational overheads — over 1 Peta FLOP/s per inference[2], becoming a major bottleneck for deployment in latency-sensitive and resource-constrained environments. Additionally, model compression techniques such as sparsification and quantization, although beneficial, often suffer from a lack of support by conventional hardware like GPUs, particularly when dealing with unstructured sparsity which, while preserving algorithmic accuracy, fails to translate into real-world performance gains.\par
\hspace{2em}Considering the limitations of current hardware for LLM, Field Programmable Gate Arrays (FPGAs) stand out as a particularly effective solution. FPGAs offer nuanced advantages over GPUs, including flexible hardware customization that can better accommodate the unique computational paradigms of LLMs such as varying sparsity patterns and mixed-precision quantization. The reconfigurability of FPGAs allows for the tuning of hardware algorithms to optimize both computational throughput and memory utilization, which is critical in LLM operations.\par
\section{SYSTEM DESIGN} 
\subsection{SpeedLLM Architecture} 
This paper proposes the SpeedLLM, an innovative acceleration solution implemented on the Xilinx Alveo U280 FPGA in Fig.1, specifically tailored for efficient inference of Tinyllama. We introduce key innovations catering to optimizing neural network computations specifically engineered for higher efficiency operations on FPGA platforms. Each of these innovations addresses critical inefficiencies in traditional FPGA neural network implementations.The main contributions of this research are threefold:\par
\begin{itemize}
    \item Customized data pipeline: We propose a multi-level read-compute-write iteration that minimizes the iterative and time-consuming cycles, obtaining an increase in the throughput and a reduction in the execution time by ensuring that compute units are constantly fed with data, avoiding idle times.
\end{itemize}
\begin{itemize}
    \item Memory Allocation Reuse Strategy: This strategy implements a cyclic or loop-back use of memory where each segment is reused after data processing is complete, without waiting for all processing to conclude. This cyclic reuse is managed through efficient scheduling algorithms that track memory usage patterns and predict availability, thus facilitating a more continuous and seamless data feed into the processor.
\end{itemize}
\begin{itemize}
    \item Operators Fusion of Llama2: Fusing operations into a single, composite operator minimizes the intermediate data writes/read between operations, reducing the processing time and memory usage.
\end{itemize}

\section{EXPERIMENTS AND RESULTS}
\subsection{Evaluation Setup}
We use a Llama2 architecture model series trained on the TinyStories dataset, intended for use in the llama2.c project. We use the stories 15M dataset in Tinyllama and tokenizer.bin in llama2.cpp. We implement the accelerator on the real system with U280 FPGAs, verified with RTL emulation using Vitis 2021.1.
\subsection{Evaluation Results}
\subsubsection{Latency \& Thoughput.}Latency measures the total time taken for complete inference by the timing function in the host program, while throughput quantifies the decoding speed by calculating the ratio of output tokens to the duration of the decode stage. Fig.2(a) shows  that our accelerator significantly surpasses the unoptimized accelerator, delivering a latency speedup of up to 4.8 times. 
\subsubsection{Energy efficiency.}We further evaluate energy efficiency. Fig.2(b) shows the energy efficiency of our accelerator, none parallel tech. one, and none fused one. Compared to no fuse accelerator, our method achieves 1.01× energy efficiency, mainly due to reduced redundant off-chip memory communications through the llama model. With higher throughput and comparable poweruse, ours achieves 1.18× better energy efficiency than an unoptimized accelerator. In terms of cost efficiency (Tokens per second per dollar), GPUs typically cost more than FPGAs. The V100S, A100, and Alveo U280 are priced around \$12,000, \$17,000, and \$8,000 respectively[3]. As a result, SpeedLLM on the U280 demonstrates superior average cost effectiveness.
\begin{figure}[htbp]
	\centering
	\begin{subfigure}{0.45\linewidth}
		\centering
		\includegraphics[width=0.9\linewidth]{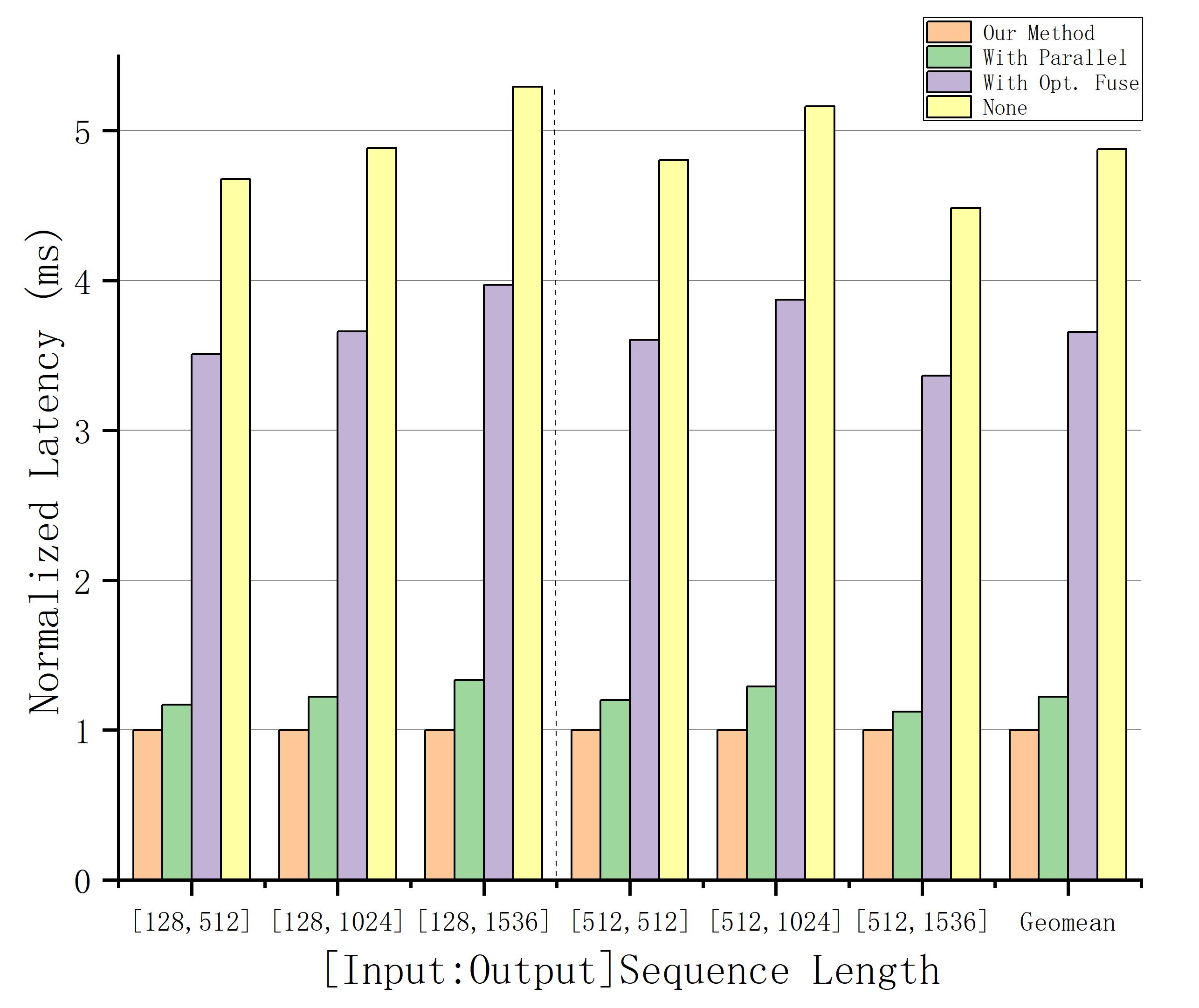}
		\caption{Normalized Latency}
		\label{chutian3}%文中引用该图片代号
	\end{subfigure}
	\centering
	\begin{subfigure}{0.45\linewidth}
		\centering
		\includegraphics[width=0.9\linewidth]{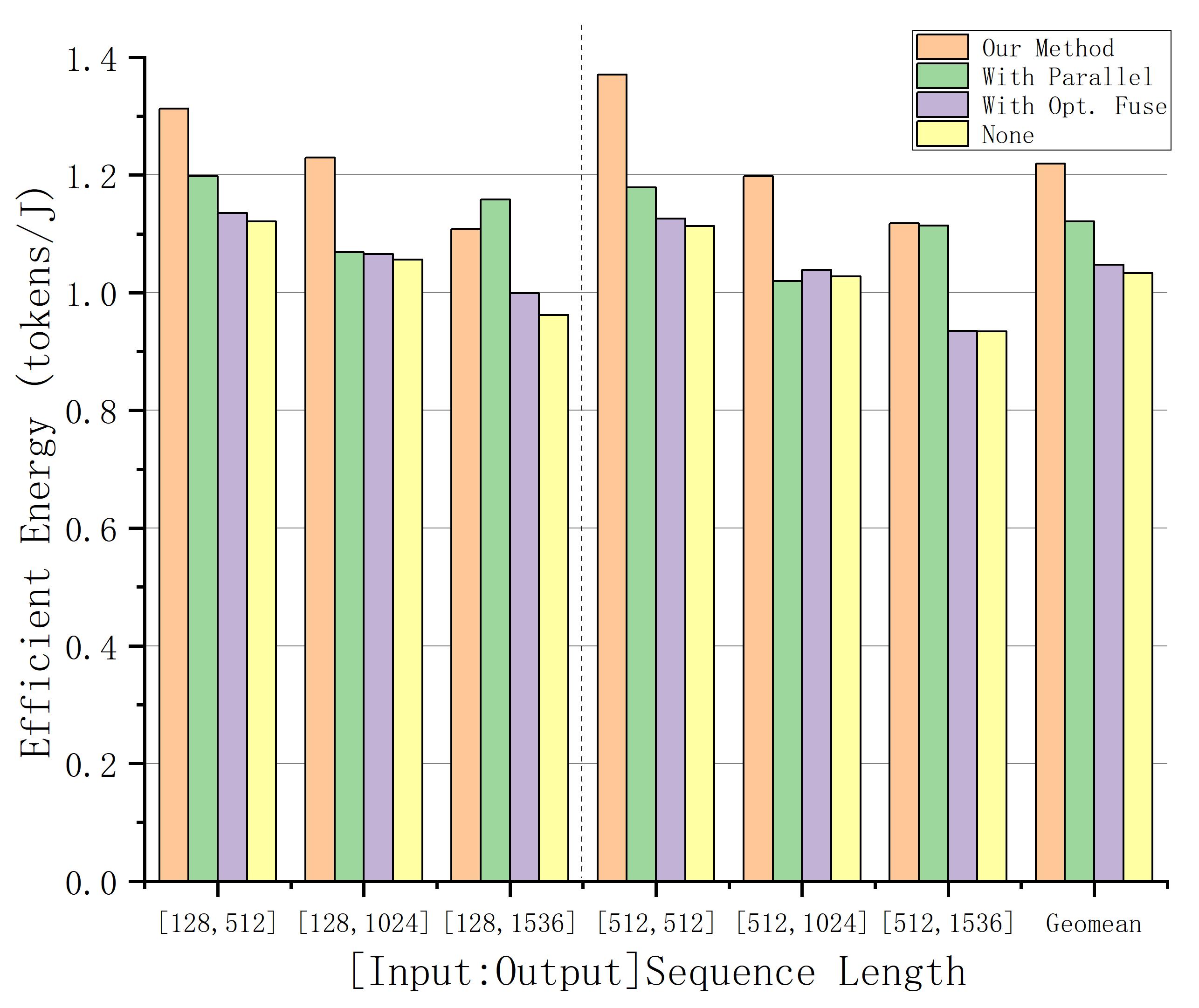}
		\caption{Effective energy}
		\label{chutian3}%文中引用该图片代号
	\end{subfigure}
	\caption{The performance of SpeedLLM}
	\label{da_chutian}
\end{figure}
\section{CONCLUSIONS}
The SpeedLLM builds upon and extends the existing research landscape by integrating several proven optimization strategies into a single coherent system that functions efficiently on the U280 FPGA platform. By implementing effective methods on LLMs and hardware design, our accelerator significantly enhances the performance capabilities of computing devices, driving forward the potential for real-world applications of deep learning in resource-constrained environments.\par

\bibliographystyle{ACM-Reference-Format}
\bibliography{sample-base}

[1]Hongzheng Chen, et al. 2024. Understanding the Potential of FPGA-based Spatial Acceleration for Large Language Model Inference.ACM Trans.Reconfigurable Tech.Syst.18,1,Article 5,29 pages.\par
[2]Nazanin Farahpour, et al. 2020. FPGA-based Near Data Processing Platform Selection Using Fast Performance Modeling. In The 21st ACM SIGPLAN/SIGBED Conference on Languages, Compilers, and Tools for Embedded Systems (LCTES '20).\par
[3]Shulin Zeng, et al. 2024. FlightLLM: Efficient Large Language Model Inference with a Complete Mapping Flow on FPGAs. In Proceedings of the 2024 ACM/SIGDA International Symposium on Field Programmable Gate Arrays (FPGA '24).\par


\end{sloppypar}
\end{document}